\input{aipcheck}

\documentclass[
    ,final            
   ,numberedheadings 
  ]
  {aipproc}

\layoutstyle{6x9}
\usepackage{epsfig} 
\usepackage{graphicx}
\usepackage{amssymb}
\usepackage{natbib}
\usepackage{times}
\usepackage{wrapfig}
\def\int{{\em INTEGRAL}}
\def\XMM{{\em XMM-Newton}}

\def\swift{{\em Swift}}

\def\j16479{IGR\,J16479-4514}

\def\igrj11{IGR\,J11215-5952}

\def\2s{2S\,0114+65}
 
\begin{document}

\title{The first X-ray eclipse of IGR\,J16479-4514?}


\author{E. Bozzo}{
  address={INAF - Osservatorio Astronomico di Roma, Via Frascati 33, 00044 Rome, Italy.}
}

\author{L. Stella}{
  address={INAF - Osservatorio Astronomico di Roma, Via Frascati 33, 00044 Rome, Italy.}
}

\author{G. Israel}{
  address={INAF - Osservatorio Astronomico di Roma, Via Frascati 33, 00044 Rome, Italy.}
}

\author{M. Falanga}{
  address={CEA Saclay, DSM/IRFU/Service d'Astrophysique, F-91191, Gif sur Yvette, France.} 
}

\author{S. Campana}{
  address={INAF - Osservatorio Astronomico di Brera, via Emilio Bianchi 46, I-23807 Merate (LC), Italy.} 
}

\begin{abstract} 
We report on the first long ($\sim$32~ks) pointed \XMM\ observation of the supergiant fast X-ray transient 
\j16479.\   
Results from the timing, spectral and spatial analysis of this observation show 
that the X-ray source \j16479\ underwent an episode of sudden obscuration, possibly an 
X-ray eclipse by the supergiant companion.   
We also found evidence for a soft X-ray extended halo around the source that is most readily 
interpreted as due to scattering by dust along the line of sight to \j16479.\ 
\end{abstract}

\maketitle

\section{Introduction}
\label{sec:intro}

Supergiant Fast X-ray transients (SFXTs) are a recently discovered subclass of 
high mass X-ray binaries. These sources display sporadic outbursts lasting from minutes 
to hours with peak luminosities of $\sim$10$^{36}$-10$^{37}$~erg~s$^{-1}$, 
and spend long time intervals at lower X-ray luminosities, 
ranging from $\sim$10$^{34}$~erg~s$^{-1}$ to 
$\sim$10$^{32}$~erg~s$^{-1}$. 
Proposed models to interpret the SFXT behavior generally involve accretion onto a neutron star 
(NS) immersed in the clumpy wind of its supergiant companion \citep{zand05}. 
The X-ray luminosity during lower activity states is likely due to    
accretion onto the NS at much reduced rate than in outburst \citep{walter07,sidoli07,bozzo08} . 
Here we report on the first long, high sensitivity observation ($\sim$32~ks) of the SFXT 
\j16479.\ This observation was carried out shortly after \swift\ discovered   
a very bright outburst from this source \citep{romano08b}, and was aimed at  
investigating the low level emission of this source and gaining insight in the physical 
mechanisms that drive this activity. 

\section{Data analysis and results}
\label{sec:observation}

\begin{figure}
\tiny
\centerline{\hbox{
\includegraphics[scale=0.25,angle=-90]{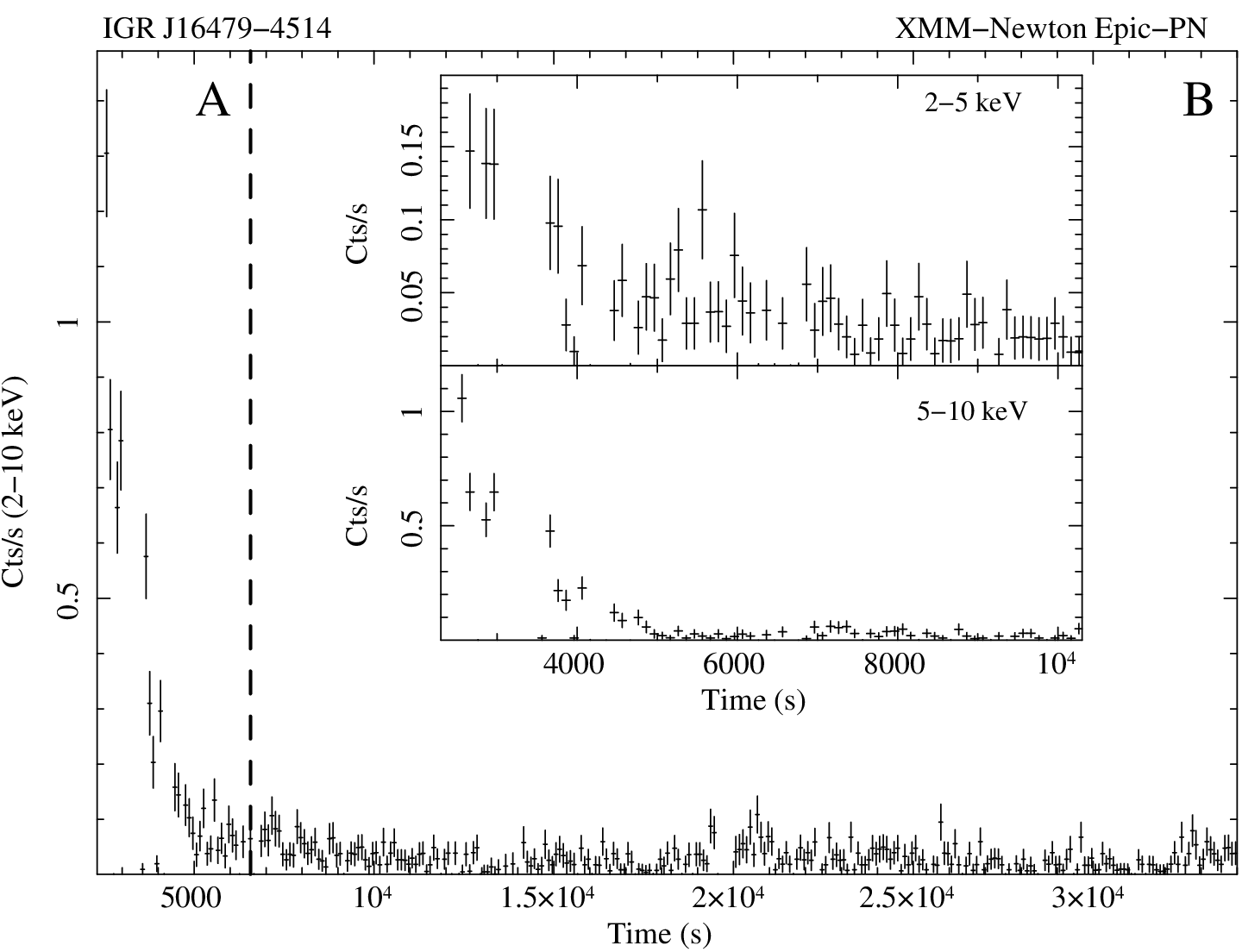}
\hspace{0.3cm}
\includegraphics[scale=0.25,angle=-90]{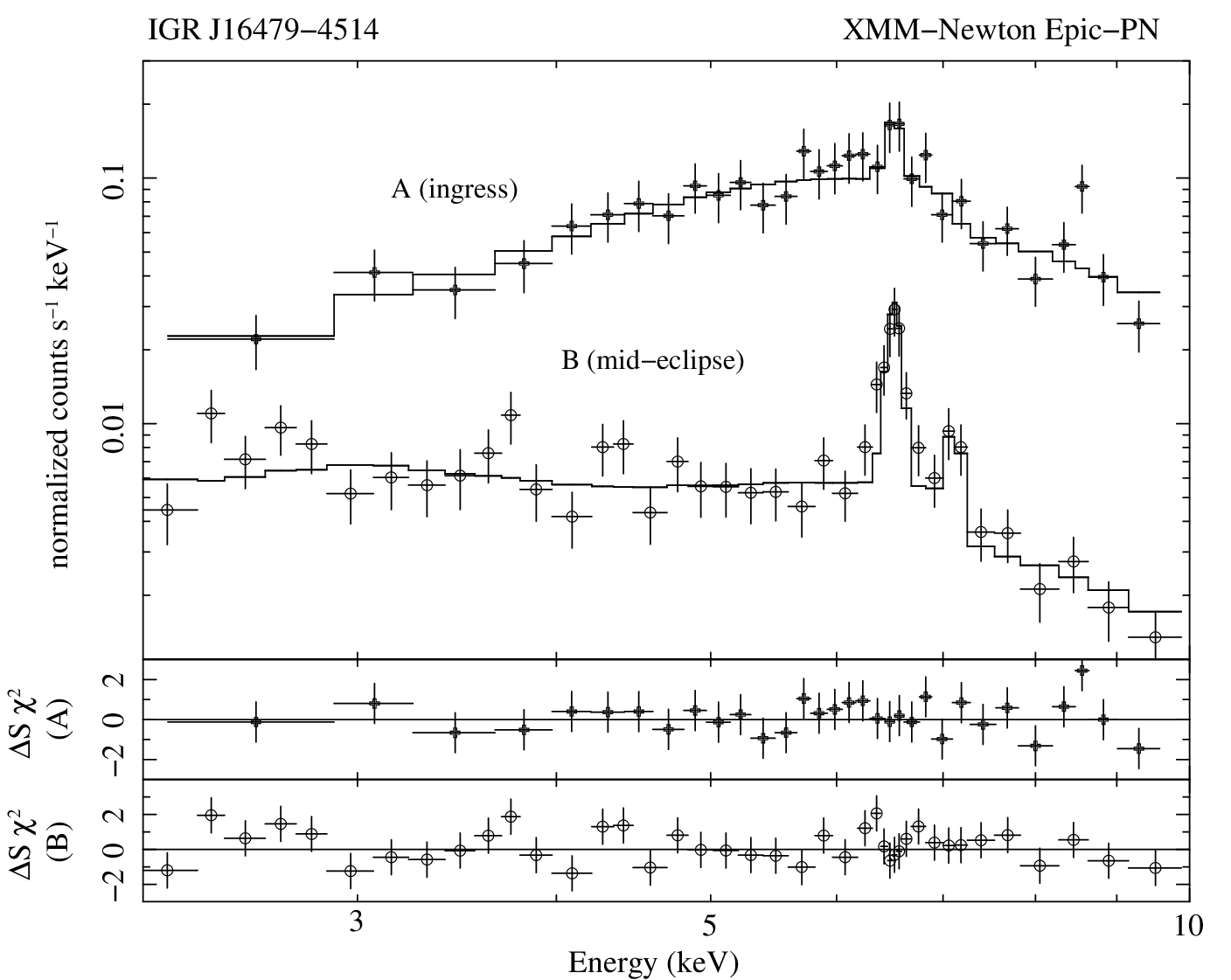}}}
\caption{{\it Left panel}: EPIC-PN light curve of \j16479\ in  the 2-10 keV band. The bin time is 100 s 
and the start time is March 21 16:42:42UT. {\it Right panel}: 2-10~keV spectra extracted from intervals A  
and B. The best fit models and the residuals from these fits are also shown. 
The best fit parameters for state A ($\chi^2$/d.o.f. = 20.4/26) are: 
$N_{H,\alpha}$=35$^{18}_{-13}$$\times$10$^{22}$~cm$^{-2}$, 
$N_{H,\beta}$=9$^{-9}_{6}$$\times$10$^{22}$~cm$^{-2}$, 
$I_{\alpha}$=1.8$^{0.4}_{-0.3}$$\times$10$^{-3}$, 
$I_{\beta}$=5$^{8}_{-4}$$\times$10$^{-3}$, $E_{\rm ln1}$=6.53$^{0.06}_{-0.07}$~keV, 
$I_{\rm ln1}$=5$^{3}_{-3}$$\times$10$^{-5}$, $EW_{\rm ln1}$=0.15~keV. 
The measured X-ray flux in this state is $F_{\rm 2-10~keV}$=10$^{-11}$~erg/cm$^{2}$/s, 
corresponding to $L_{\rm X}$=2.9$\times$10$^{34}$~erg/s  
\citep[the source distance is 4.9~kpc;][]{rahoui08}. 
For the B state ($\chi^2$/d.o.f. =36.5/33) we found: 
$N_{H,\alpha}$=54$^{26}_{-25}$$\times$10$^{22}$~cm$^{-2}$, 
$I_{\alpha}$=1.1$^{0.5}_{-0.4}$$\times$10$^{-4}$, $I_{\beta}$=8$^{2}_{-3}$$\times$10$^{-4}$, 
$E_{\rm ln1}$=6.51$^{0.03}_{-0.02}$~keV, $I_{\rm ln1}$=1.8$^{1.0}_{-0.4}$$\times$10$^{-5}$,  
$EW_{\rm ln1}$=0.77~keV, $E_{\rm ln2}$=7.11$^{0.06}_{-0.09}$~keV, 
$I_{\rm ln2}$=8$^{7}_{-4}$$\times$10$^{-6}$, $EW_{\rm ln2}$=0.28~keV, 
$F_{\rm 2-10~keV}$=7.5$\times$10$^{-13}$~erg/cm$^{2}$/s  
($L_{\rm X}$=2.2$\times$10$^{33}$~erg/s). 
Here $F_{\rm 2-10~keV}$ is the absorbed flux in the 2-10~keV band, and 
errors are at 90\% confidence level. } 
\label{fig:lcurve} 
\end{figure}

\XMM\ \citep{j01} observed \j16479\ between March 21 14:40:00UT and March 22 01:30:00UT.  
We show In figure~\ref{fig:lcurve} the 2-10~keV EPIC-PN light curve and spectra 
of the observation.   
In the first $\sim$4~ks \j16479\ was caught during the decay from a higher (``A'')
to a lower (``B'') flux state.  
The shape of the above light curves and evolution of 
the spectrum across the state A-state B transition 
presented remarkable similarities with the eclipse ingress 
of eclipsing X-ray sources such as OAO\,1657-415 \citep{audley06}. 
Moreover the slower decay of the 
soft X-ray light curve (in turn similar to that observed in OAO\,1657-415)
suggested that \j16479\ is seen through an extended 
dust-scattering halo \citep{day91}. To confirm this we analysed the radial distribution 
of the X-ray photons detected from \j16479\ and compared it with the
point spread function (PSF) of the \XMM\ telescope/EPIC-PN camera.  
We found that a 30'' extended dust scattering halo located halfway between us and \j16479\ 
agrees well with the e-folding decay time of the soft X-ray light curves. 
Motivated by the above findings  
we considered a spectral model that has been used in studies
of eclipsing X-ray binaries seen through a dust-scattering halo, that is:  
$I(E)$=$e^{\sigma(E) N_{\rm H,\beta}}$$I_{\beta}$$E^{-\beta}$+$e^{\sigma(E) N_{H,\alpha}}$[$I_{\alpha}$$E^{-\alpha}$+   
$I_{\rm ln1}$$e^{-(E-E_{\rm ln1})^2/(2\sigma_{\rm ln1}^2)}$+$I_{\rm ln2}$$e^{-(E-E_{\rm ln2})^2/(2\sigma_{\rm ln2}^2)}$].   
In practice we fitted a model with two continuum components, a single power law 
of slope $\alpha$ (with normalisation $I_{\alpha}$ and column density $N_{H,\alpha}$) 
and a power law of slope $\beta$ (with normalisation $I_{\beta}$ and column density $N_{H,\beta}$). 
The former component was taken to approximate the sum of two $\alpha$-slope power laws: the first  
represents the direct component 
dominating before the eclipse ingress (i.e. the component due to the accretion process onto the NS), whereas  
the second represents the wind scattered component dominating during the eclipse \citep{audley06}. 
The second power law component, with photon index $\beta$=$\alpha$+2, originates from 
small angle scattering of (mainly) direct photons off interstellar dust grains 
along the line of sight \citep{day91}. The Gaussians in the above equation represent 
iron features around $\sim$6.4~keV and $\sim$7.0~keV, arising from the 
reprocessing of the source radiation in the supergiant wind.  
We kept $\alpha$ fixed  at 0.98, and thus $\beta$=2.98 \citep{romano08}. The best fit parameters 
are reported in the caption of Fig.~\ref{fig:lcurve}.  
\begin{figure}
\centering
\includegraphics[scale=0.3,angle=-90]{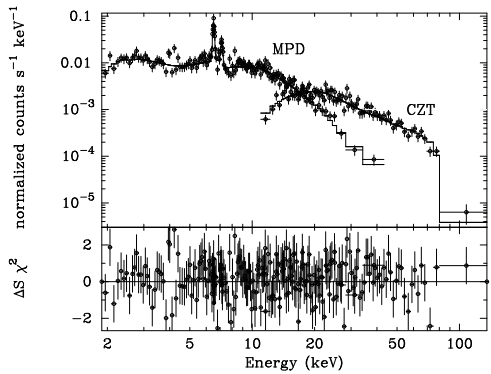}
\caption{Simulated Simbol-X spectrum of \j16479 during state B. Both the results for 
the MPD (Macro-Pixel Detector), and the CZT(Cd(Zn)Te detector) are shown.} 
\label{fig:simbolx} 
\end{figure}

In state A the relatively high source flux and small EW of the iron fluorescence line 
testifies that (most of) the emission is likely due to the direct component. 
In state B the ratio $I_{\alpha}$/$I_{\beta}$ is larger than the corresponding value  
obtained during state A, whereas the EW of the Fe-line at $\sim 6.5$~keV 
increased from $\sim$150~eV to $\sim$770~eV. 
We also found evidence ($\sim$2$\sigma$) for an additional Fe-line at 
$\sim 7.1$~keV with a $\sim$300~eV EW, consistent with being the $K_{\beta}$. 
The most natural interpretation of this is that in state B  
the direct emission component is occulted along our line of sight, while the spectrum 
we observe is the sum of a dust scattered component, dominating at lower energies, and a 
wind scattered component characterised by a high absorption. 
The marked increase in the EW of the Fe-line at $\sim$6.5~keV across the state A-state B 
transition testifies that the region where the line is emitted is larger 
than the occulting body (the supergiant companion, if we are dealing with an eclipse); this also provides further evidence that 
the uneclipsed emission in state B at hard X-ray energies (the $\alpha$-slope power law) arises mostly 
from photons scattered by the wind in the immediate surrounding of the source.  

We conclude that \j16479\ is the first SFXT 
that displayed evidence for an X-ray eclipse. 
Further observations of \j16479\ and other SFXTs in quiescence, will improve our knowledge 
of the low level emission of these sources, and will help clarifying if  
X-ray eclipses are common in SFXTs. In Fig.~\ref{fig:simbolx} we show 
the X-ray spectrum of \j16479\ during state B that would be observed by Simbol-X. 
We used an exposure time of 28~ks and the latest Simbol-X simulation tool 
available (see http://www.apc.univ-paris7.fr/Simbolx2008/). \\ 

\small {\it We thank Norbert Schartel and the XMM-Newton staff, 
for carrying out this ToO observation; EB thanks L. Sidoli and P. Romano 
for their collaboration during the early stages of this work.}  

\bibliographystyle{aipproc}   

\end{document}